\documentclass[a4paper,11pt]{article}

\usepackage{jinstpub}
\usepackage{lineno}

\usepackage[utf8]{inputenc}

\usepackage{hyperref} 
\usepackage{url}
\usepackage{graphicx}

\usepackage{multirow}

\usepackage{graphics}
\usepackage{natbib}
\graphicspath{{./}}

\usepackage{xcolor}
\usepackage{textcomp}

\def\unit#1{\hbox{$\,{\rm #1}$}}

\usepackage{orcidlink}

\title{The EXTRA-BL4S experiment for the measurement of the energy and angular distributions of transition radiation X-rays}

\author[a,1]{M.~N.~Mazziotta\,\orcidlink{0000-0001-9325-4672}\note{Corresponding authors.},}
\author[a,b,1]{F.~Loparco\,\orcidlink{0000-0002-1173-5673},}

\author[c]{A.~Anelli,}
\author[c]{M.~M.~Belviso,}
\author[c]{A.~Buquicchio,}
\author[c]{E.~V.~Cassano,}
\author[c]{M.~De~Cosmo,}
\author[c]{P.~Ginefra,}
\author[c]{M.~L.~Martulli,}
\author[c]{C.~Picci,}
\author[c]{D.~Picicci,}
\author[c]{R.~D.~Soriano,}
\author[c]{A.~P.~Tatulli,}
\author[c]{G.~Tripaldella,}
\author[c]{V.~M.~Zupo,}
\author[c]{M.~F.~Muscarella,}
\author[c]{S.~Turbacci,}

\author[d]{M.~Boselli\,\orcidlink{0000-0003-3540-8228},}
\author[d,2]{C.~B.~da~Cruz~E~Silva\,\orcidlink{0000-0002-1231-3819},\note{Now at LIP - Laboratório de Instrumentação e Física Experimental de Partículas Avenida Prof. Gama Pinto 2, Complexo Interdisciplinar (3is), 1649-003 Lisboa, Portugal}}
\author[d]{M.~Joos\,\orcidlink{0000-0001-8184-5598}}
\author[e]{and P.~Sch\"utze\,\orcidlink{0000-0003-4802-6990}}

\affiliation[a]{Istituto Nazionale di Fisica Nucleare, Sezione di Bari, \\ via Orabona 4, I-70126 Bari, Italy}
\affiliation[b]{Dipartimento di Fisica dell'Universit\`a e del Politecnico di Bari, \\ via Amendola 173, I-70126 Bari, Italy}

\affiliation[c]{\textbf{The EXTRA Team} Liceo Scientifico Statale "A. Scacchi",  \\ Corso Cavour 241, I-70121 Bari, Italy}
\affiliation[d]{CERN, the European Organization for Nuclear Research, \\ Esplanade des Particules 1, 1211 Geneva, Switzerland}
\affiliation[e]{Deutsches Elektronen-Synchrotron DESY, \\ Notkestr. 85, 22607 Hamburg,
Germany}

\emailAdd{mazziotta@ba.infn.it}
\emailAdd{francesco.loparco@ba.infn.it}


\arxivnumber{2301.11247} 

\abstract{
We have designed and implemented an experiment to measure the angular distributions and the energy spectra of the transition radiation X-rays emitted by fast electrons and positrons crossing different radiators. Our experiment was selected among the proposals of the 2021 Beamline for Schools contest, a competition for high-school students organized every year by CERN and DESY, and was performed at the DESY II Test Beam facility area TB21, using a high-purity beam of electrons or positrons with momenta in the range from $1$ to $6 \unit{GeV/c}$. The measurements were performed using a $100\unit{\mu m}$ thick silicon pixel detector, with a pitch of $55\unit{\mu m}$. Our results are consistent with the expectations from the theoretical models describing the production of transition radiation in multilayer regular radiators.
}

\begin{document}
\maketitle
\flushbottom

\section{Introduction}
\label{sec:intro}

In recent years, high-school physics curricula increasingly include topics related to modern high-energy physics and particle detectors. Universities and research centers promote several programs to bring high-school students in touch with modern physics and the scientific research. The Liceo Scientifico ``A.~Scacchi'' in Bari has taken part in such projects for years, and in 2021 the school promoted the participation of a team of students of the $12^{th}$ and $13^{th}$ grade in the Beamline for Schools (BL4S) competition. 

BL4S is organized by CERN in collaboration with DESY, and offers to groups of high-school students the unique opportunity to propose a scientific experiment at a particle accelerator facility and to win a trip to perform it. Because of the maintenance of CERN accelerators, the experiment was performed at the DESY II Test Beam facility in Hamburg. The students, coordinated by their physics teachers and under the supervision of experienced researchers from the Physics Department of the Bari University and from the INFN Unit in Bari, won the competition. 

The goal of the experiment conceived by the team was to study the transition radiation emitted by fast electrons and positrons crossing different kinds of radiators. This paper provides a short presentation of the BL4S competition and presents the experiment and the result obtained by the team during their beam time in Hamburg in September 2021.

\section{The BL4S competition}
\label{sec:bl4s}

Beamline for Schools (BL4S)~\cite{BL4S} is a physics competition organised by CERN and DESY, which invites high-school students from all over the world to propose an experiment to be performed at a particle accelerator. Each team has to write an original scientific proposal, explaining the theoretical background of the selected topic, and describing both the procedure to carry it out at a test beam facility and the results that they expect to find. A jury of experts, including scientists of CERN and DESY, review the proposal and select two teams (three from 2022 on) that win a trip to a fully equipped beam line of a particle accelerator. 

From 2014 to 2018 the winning experiments took place at the test beam area of the CERN Proton Synchrotron (PS) accelerator. In 2019 the competition moved to the DESY II Test Beam Facility (Hamburg, Germany)~\cite{DESYIITB}. The partnership between CERN and DESY allowed BL4S to continue during the three-year long shutdown of the CERN accelerator complex for upgrade and maintenance.  

The competition is structured in several preparatory phases, which include conferences and meetings with the organisers. Once the competition is announced, usually in Autumn, interested teams start preparing their proposals. Teams can include students either from the same school or from different schools. Having teams representing two schools or more is not unusual. During the proposal preparation, students are involved in an intense research project. After the conception and design of their experiment, the participants must write a well structured proposal and submit it on time. The students are not alone in this process, but they are guided by their coaches, who provide them with details on particle physics and teach them the necessary technical skills. Team coaches can be teachers, parents or scientists of local universities. It is important that students are well aware of each scientific detail of the proposed experiment, so that the theoretical background is clear and solid. The students are required to write down in detail how they intend to use the particle beam for their measurements and which equipment and detectors they need. Moreover, participants often complement their theoretical hypothesis with computer simulations. In fact, it is fundamental that students acquire the rudimentary programming skills that will be required in case of victory. Lastly, conclusions must contain the team’s expectations and motivation, which play a significant role in the jury’s decision. The BL4S organisers are always available to answer questions that the teams might have during the preparation of their proposals. Many teams contact them to discuss the feasibility of their experiments or practical problems that they encounter.

In the final phase of the competition, a jury consisting of more than 50 volunteers selects the teams that are invited to a research institute to perform their experiment together with support scientists. Prior to the visit, the winning teams work remotely with the BL4S scientists to refine their experiments and perform a detailed planning.

The beam time of the winning teams usually happens just after the summer, and the students have 12 full days of access to the experimental area to perform their measurements, supervised by the support scientists. During their stay, they work as a team of professional scientists would do and they complement their scientific experience with visits and lectures.

After taking the data at the beam line, the teams are encouraged to analyse their data to answer the scientific question of the initial proposal, and to write a paper. During this phase, the team members stay in close contact with the BL4S support scientists and the team coaches.

\section{The EXTRA experiment}
\label{sec:setup}

The EXTRA (Electron X-ray Transition RAdiation) experiment is designed to study the transition radiation (TR)~\cite{Ginzburg:1945zz} emitted by fast electrons and positrons crossing different radiators. 

Highly relativistic particles crossing the boundary between materials with different dielectric constants can produce TR in the X-ray region. However, since the yield of TR photons emitted at a single interface is considerably small (it is of the order of the fine structure constant $\alpha \approx 1/137$), multiple boundaries are needed to enhance the X-ray production. Periodic radiators, consisting of stacks of thin foils of dielectric material separated by thicker air gaps, are commonly used in transition radiation detectors (TRDs)~\cite{Artru:1975xp}.

The main features of the TR emitted by a periodic radiator depend on the kinematic properties of the radiating particles and on the radiator properties. They can be summarized as follows~\cite{Zyla:2020zbs}:
\begin{enumerate}
\item The effective TR photon emission starts at a threshold Lorentz factor, which is given by $\gamma_{thr}=d_{1}\omega_{1}/c$, where $d_{1}$ is the thickness of the foils, while $\omega_{1}$ is the plasma frequency of the foil material.
\item The TR emission increases with the Lorentz factor $\gamma$ until it reaches saturation at
$\gamma_{sat}=\gamma_{thr} \sqrt{d_{2}/d_{1}}$,
where $d_{2}$ is the thickness of the air gaps.
\item Most of the TR energy is emitted near the energy $\hbar \omega_{max}=\hbar \omega_{1}^{2}d_{1}/2\pi c$.
\item The angular distribution of TR photons exhibits a few maxima and extends up to $\theta_{max}=\sqrt{1/\gamma^{2}+\omega_{1}^{2}/\omega^{2}}$.
\end{enumerate}

\begin{figure}[!b]
	\begin{center}
	\includegraphics[bb=50 70 400 200,scale=0.9]{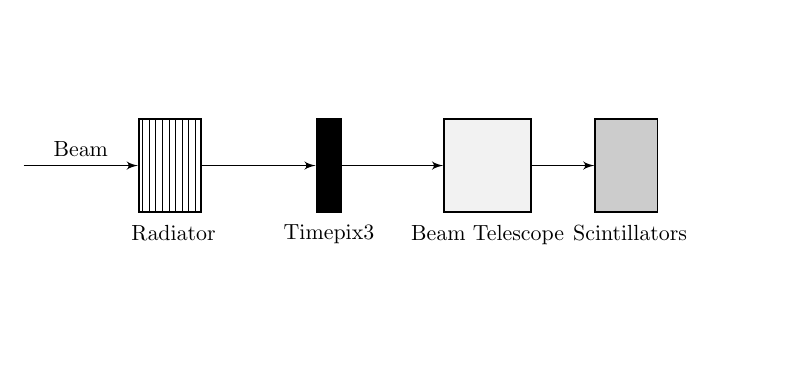}
	\end{center}
	\caption{Schematic view of the experimental setup.}
	\label{fig:setup} 
\end{figure}

\begin{figure}[!t]
	\begin{center}
	\includegraphics[width=0.9\textwidth,height=0.60\textheight]{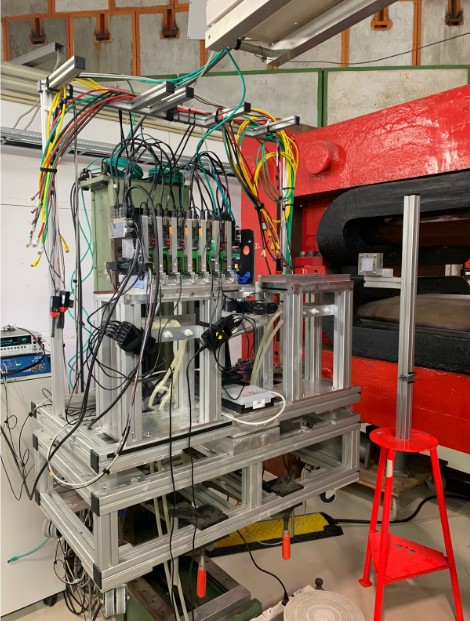}
	\includegraphics[width=0.9\textwidth,height=0.30\textheight]{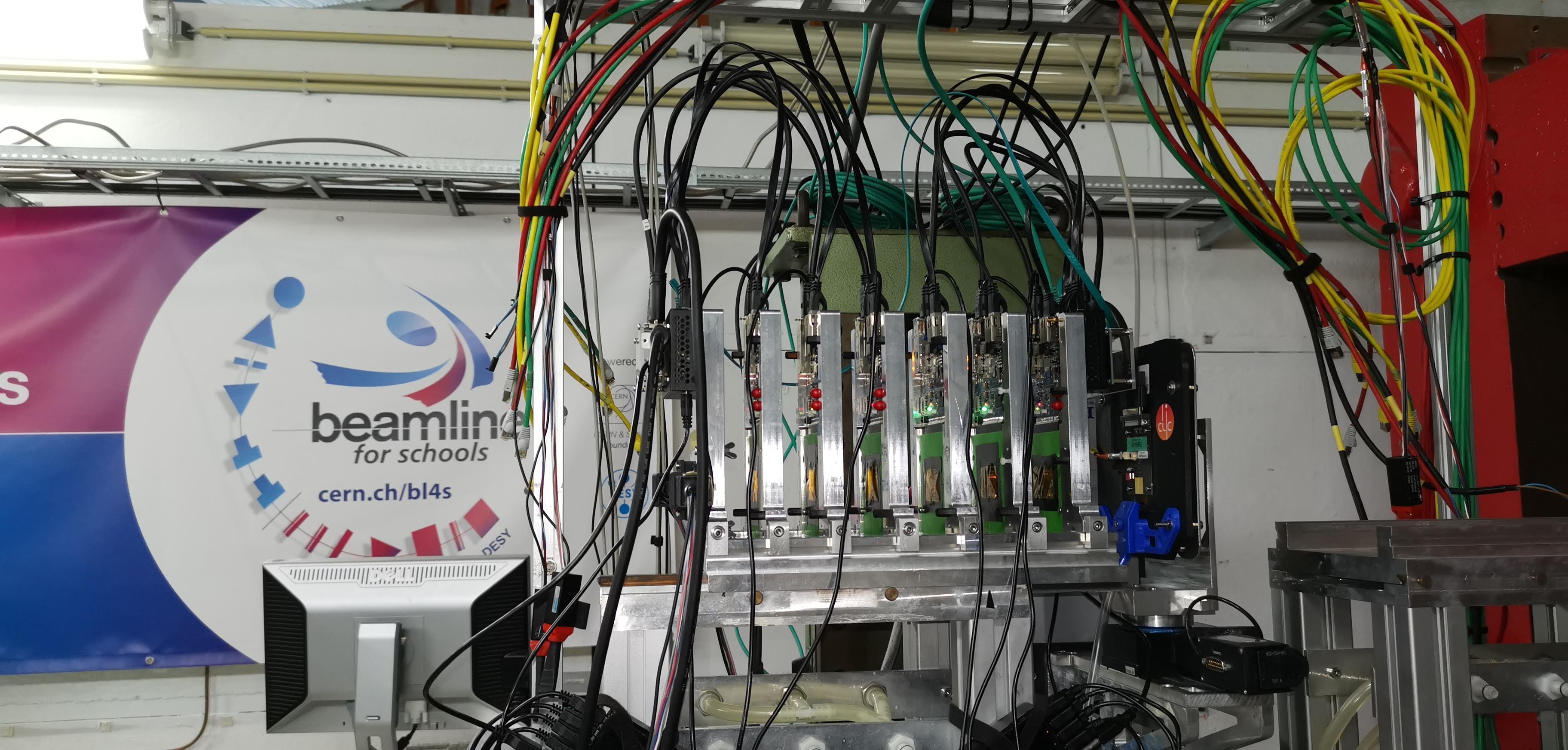}
	\end{center}
	\caption{Pictures of the experimental setup.}
	\label{fig:setup1} 
\end{figure}

We have designed an experimental setup to measure the energy spectra and the angular distributions of the TR X-rays emitted by fast electrons and positrons crossing different radiators. Similar measurements were performed in the past at the CERN SPS with beams of $20 \unit{GeV/c}$ electrons  
and of $120$, $180$ and $290 \unit{GeV/c}$ muons, using silicon strip detectors~\cite{Alozy:2019kiq}, silicon pixel detectors~\cite{Alozy:2020wsu,Dachs:2020lga} and GaAs pixel detectors~\cite{Alozy:2020ges,Dachs:2020lga}. Parallel to the measurements, an effort to develop accurate Monte Carlo simulations of the TR process is being carried out~\cite{Savchenko:2020eos}. One of the goals of these activities is that of exploiting TR for the identification of charged hadrons in the $\unit{TeV}$ energy region~\cite{Albrow:2018kxz}. In this region all hadrons have Lorentz factor exceeding the typical threshold values for TR production (usually $\gamma_{thr} \sim 500 \div 1000$), and the simultaneous measurement of the energies and of the emission angles of TR X-rays can help to discriminate among different hadron species.

Our measurements were performed at the DESY II Test Beam Facility~\cite{DESYIITB} area TB21, using a beam of either electrons or positrons with momenta in the range from $1$ to $6 \unit{GeV/c}$. A scheme of the setup is shown in Fig.~\ref{fig:setup}, while pictures of the setup are shown in Fig.~\ref{fig:setup1}. The radiator is followed by a Timepix3 assembly containing a thin silicon pixel sensor, which is used to detect the TR X-rays. A downstream beam telescope, composed by an array of six silicon pixel detectors, is used to reconstruct the tracks of the beam particles~\cite{eudetPerformance}. A set of two trigger scintillators, located downstream of the last plane of the beam telescope, is used for triggering the data acquisition.

\begin {table}[t]
\begin{center}
\begin{tabular}{ |l| c| c| c| c|  } \hline \hline
Radiator & Foil/gap material & d$_1$ ($\mu$m) & d$_2$ ($\mu$m) & N$_f$ \\  \hline
EXTRA & polyethylene/air & 23 & 500 & 150 \\
INFN   &  polyethylene/air & 25 & 300 & 155 \\
CERN  &  polyethylene/air & 25 & 240 & 190 \\ \hline \hline
\end{tabular}
\caption{Parameters of radiators used in the beam test: $d_1$ and $d_2$ are the thickness of the foils and the gap respectively; $ N_f$ is the number of foils.  }
\label{tab:rad}
\end{center}
\end{table}

\begin {table}[t]
\begin{center}
\begin{tabular}{|l|c|c|c|}
\hline
\hline
Radiator & distance ($\unit{cm}$) & Beam particle & Beam momenta ($\unit{GeV/c}$) \\
\hline
\multirow{3}{*}{EXTRA} & $40.5$ & $e^{-}$ & $1,2,3,4,5,6$ \\
                       & $88.0$ & $e^{-}$ & $1,2,3,4,5,6$ \\
                       & $132.0$ & $e^{-}$ & $1,2,3,4,5,6$ \\
\hline
INFN & $88.9$ & $e^{-}$ & $1,2,3,4,5,6$ \\
\hline
CERN & $88.4$ & $e^{-}/e^{+}$ & $1,2,3,4,5$ \\
\hline
\hline
\end{tabular}
\caption{Summary of the data taking configurations. For each radiator the beam particle, their momenta and the distance between the radiator and the X-ray detector are reported.}
\label{tab:beam}
\end{center}
\end{table}

\begin{figure}[t!]
    \centering
    \includegraphics[width=0.95\textwidth,height=0.45\textheight]{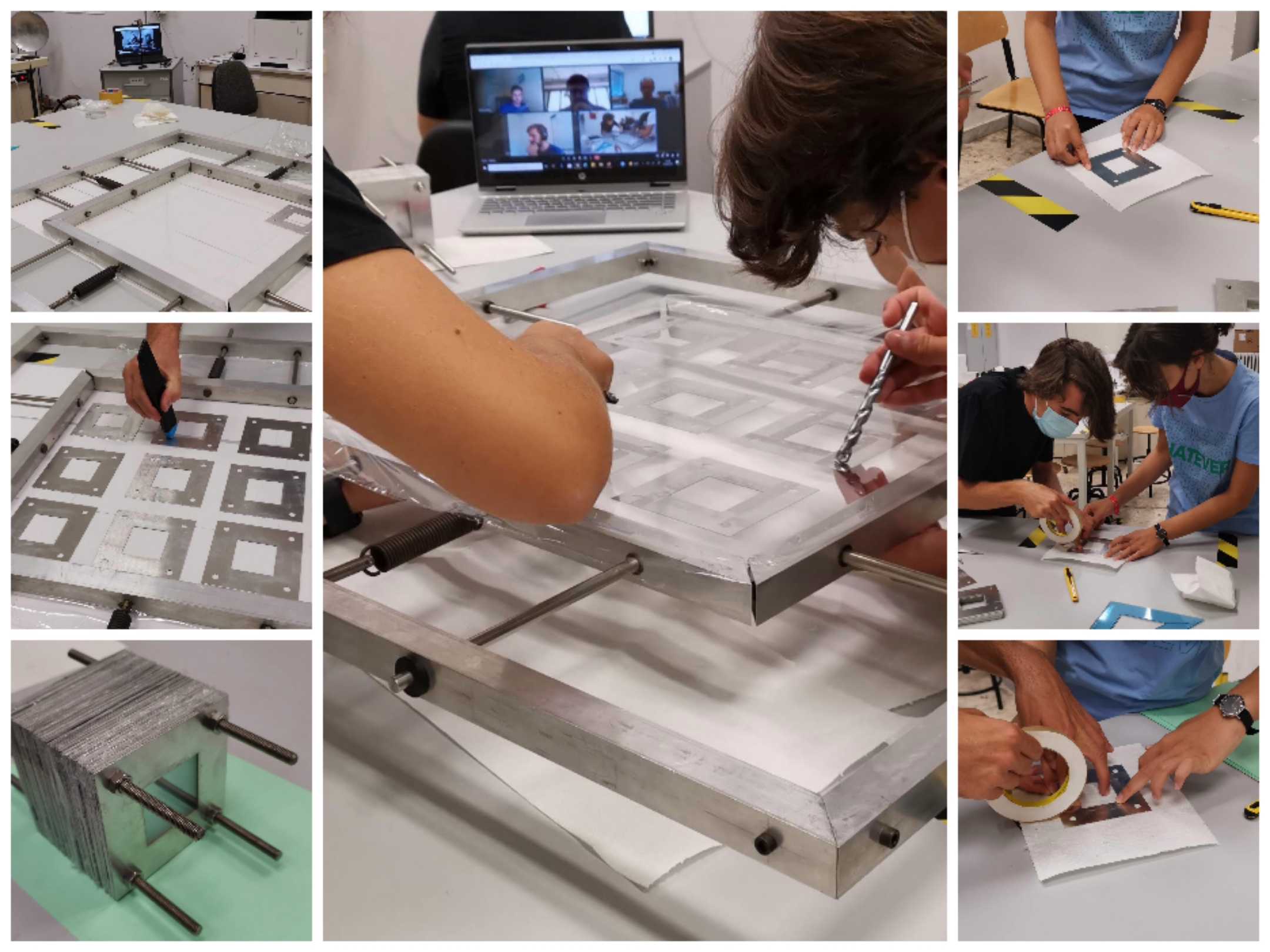}
    \caption{Assembly of the EXTRA radiator at the Liceo Scientifico "A. Scacchi".}
    \label{fig:assextrarad}
\end{figure}

In our experiment we used three different radiators, which in the following will be labelled as "EXTRA", "INFN" and "CERN" respectively. Their features are summarized in Tab.~\ref{tab:rad}. In particular, the EXTRA radiator was assembled for this measurement by the students at the Liceo Scientifico "A. Scacchi" in Bari. Fig.~\ref{fig:assextrarad} shows some picture taken during the assembly of this radiator. The INFN and CERN radiators were borrowed from the Bari INFN Group and were used in a beam test campaign performed in 2006~\cite{Brigida:2007zza}.

With these radiators, several measurements were performed, changing the beam composition and momentum, and the distance between the radiator and the X-ray detector. The different data taking configurations are summarized in Tab.~\ref{tab:beam}.

The TR X-rays were detected by a $100\unit{\mu m}$ thick silicon sensor, bump-bonded to a Timepix3 readout chip~\cite{TPX3}, consisting of a pixel matrix of $256 \times 256$ pixels with a pitch of $55\unit{\mu m}$. This silicon detector assembly was placed such that the sensor faces the radiator to mitigate prior absorption in the readout chip. The sensor of the assembly with the ID W5\_E2 was operated at a bias voltage of $-21\unit{V}$ to ensure full depletion~\cite{W5_E2}.

The pixel pitch of the silicon sensor and its distance from the radiator determine the minimum detectable angular separation $\theta_{min}$ of TR X-rays from the direction of the radiating particles, as they should be separated by at least one pixel. Its value is in fact given by $\theta_{min} \gtrsim w/d$, where $w=55\unit{\mu m}$ is the pixel pitch and $d$ is the distance of the silicon detector from the radiator. We also remark that a fraction of the TR photons emitted at angles $\theta \lesssim 3 w/d$ from the particles will be suppressed, since the particles can yield detectable signals on clusters of a few adjacent pixels. The configurations with larger distances allow to detect smaller angular separations; however, due to the X-ray absorption in the radiator and in the air gap between the radiator and the sensor, the number of detected TR X-rays will decrease with the distance from the radiator and the angular resolution will deteriorate due to multiple Coulomb scattering of the primary particles in air.

While the TR X-rays are likely absorbed by the front sensor (the photoelectric absorption probability for $10\unit{keV}$ X-rays in a $100\unit{\mu m}$ thick silicon layer is $\sim 54\%$~\cite{nist}), the radiating charged particles traverse the detector and leave an ionization track in the detectors of the beam telescope, which consists of an array of six regularly spaced silicon pixel detectors. By minimizing the space in-between the sensor planes, the scattering in air is limited, thus enabling a track resolution of a few $\unit{\mu m}$ extrapolated to the Timepix3 detector~\cite{eudetPerformance}, which is more than sufficient for an identification of the charged particle among two or more clusters in the Timepix3 detector with cluster distances larger than a pixel pitch.

Finally, the two scintillators, approximately shadowing the size of the telescope sensor planes and located at the end of the beam line, are used for triggering the data acquisition.

The data acquisition was performed using the software framework \textit{EUDAQ2}~\cite{eudaq}, which integrates the control and readout of the Timepix3 assembly and the beam telescope, and features a graphical user interface for the configuration of connected devices, starting and stopping runs and data storage. An AIDA TLU~\cite{TLU} was used to form a trigger signal as a coincidence of the signals from the two scintillators while enabling a busy-handshake with the detectors.

\section{Data analysis}
\label{sec:data}

\subsection{Conversion \& Clustering}
\label{sec:clustering}

The raw data contains a collection of hit pixels per detector plane per trigger, which defines a so-called "event", including the corresponding pixel addresses; for the data from the Timepix3 assembly, the corresponding information on the energy deposit, in form of a digitised signal, is also stored, while for the beam telescope no charge information is recorded. The collected data are converted to the \textit{ROOT TTree} format~\cite{ROOT} using the data analysis framework \textit{Corryvreckan}~\cite{corryvreckan}. In addition, this software performs a clustering procedure, which identifies adjacent hit pixels and connects them to form a so-called "cluster" under the hypothesis that pixel hits in one cluster are caused by a single incident particle. The cluster center, as an estimation on the incidence position of the particle, is calculated either as the center-of-gravity using the charge information, or as the arithmetic mean of the pixel hit positions in case of binary hit information.

The energy calibration of the silicon pixel detector is performed assuming that the most probable energy loss of $5 \unit{GeV/c}$ electrons crossing a $100\unit{\mu m}$ thick silicon layer is $25.41\unit{keV}$. This value has been calculated using a dedicated Monte Carlo simulation for the calculation of the energy losses of charged particles in thin silicon absorbers~\cite{Bichsel:1988if,Brigida:2004ff}.

\subsection{Detector alignment procedure}

The positions of the clusters in each silicon detector are evaluated in the local detector reference frame, with the $z$-axis oriented along the beam direction and the $x-y$ plane corresponding to the detector plane, with the origin in the center of the detector. In the global reference frame the $z$-axis is also directed along the beam direction, and the detectors are disposed on planes parallel to the $x-y$ plane, with their centers at the coordinates $(x_{0}^{i},y_{0}^{i},z_{0}^{i})$. Due to mechanical tolerances in the assembly of the detectors, the coordinates $(x_{0}^{i},y_{0}^{i})$ are slightly misaligned with respect to the reference values $(0,0)$. 

A dedicated alignment run has been therefore performed to evaluate the coordinates $(x^{i}_{0},y^{i}_{0})$ of the centers of the silicon detectors (the index $i=0$ refers to the Timepix3 sensor, while the indices $i=1 \ldots 6$ refer to the detectors of the beam telescope). The alignment run has been performed removing the radiator from the beam line and using $5\unit{GeV/c}$ electrons. 

We have implemented an iterative alignment procedure selecting a sample of events with only one cluster in each silicon detector. This choice is aimed to select events with only one electron track across all the detectors. In the first iteration we assume $x_{0}^{i}=0$ and $y_{0}^{i}=0$ for all detectors. We fit all the tracks with a straight line and, for each track, we evaluate the residuals in each detector as $r_{x}^{i}=x^{i}-x^{i}_{fit}$ and $r_{y}^{i}=y^{i}-y^{i}_{fit}$, where $(x^{i},y^{i})$ and $(x^{i}_{fit},y^{i}_{fit})$ are respectively the true and fitted positions of the cluster in the $i$-th detector. We then build the distributions of the residuals $r_{x}^{i}$ and $r_{y}^{i}$ and, in the next iteration, we set $x_{0}^{i}=-\mu_{x}^{i}$ and $y_{0}^{i}=-\mu_{y}^{i}$, where $\mu_{x}^{i}$ and $\mu_{y}^{i}$ are the average values of these distributions. The iterative procedure is terminated when $|\mu_{x}^{i}|<1 \unit{\mu m}$ and $|\mu_{y}^{i}|<1 \unit{\mu m}$ for all detectors. Convergence is reached after the second iteration.

\begin{figure}[!t]
\begin{center}
\includegraphics[width=0.48\textwidth,height=0.22\textheight]{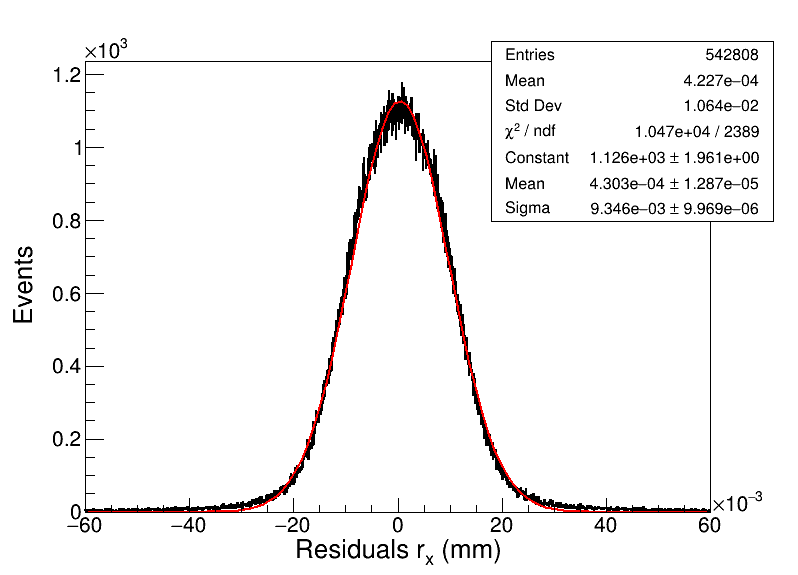}
\includegraphics[width=0.48\textwidth,height=0.22\textheight]{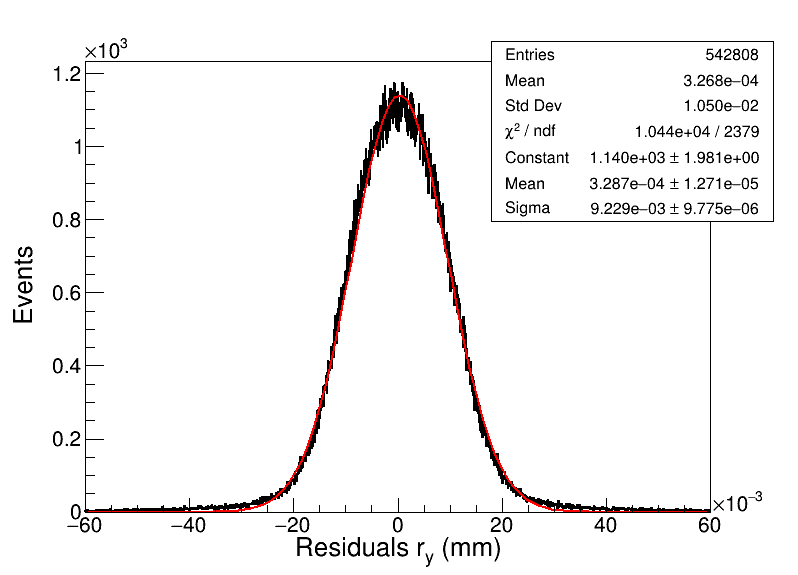}
\end{center}
\caption{Distributions of the residuals in the silicon detector equipped with the Timepix3 chip after the alignment procedure.}
\label{fig:residuals} 
\end{figure}

Fig.~\ref{fig:residuals} shows the distributions of the residuals in the silicon detector equipped with the Timepix3 chip after the alignment procedure. The RMS of the residual distributions in both the $x$ and $y$ views are of about $10\unit{\mu m}$. 

\begin{figure}[!t]
\begin{center}
\includegraphics[width=0.48\textwidth]{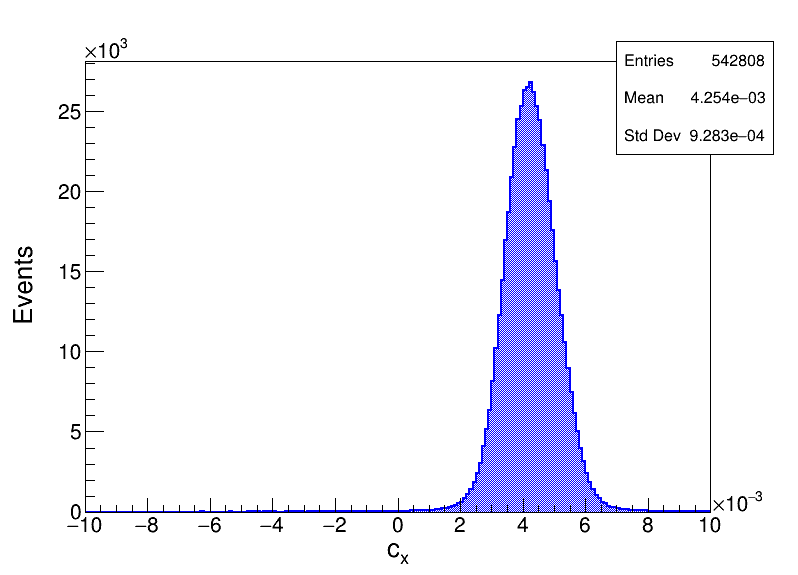}
\includegraphics[width=0.48\textwidth]{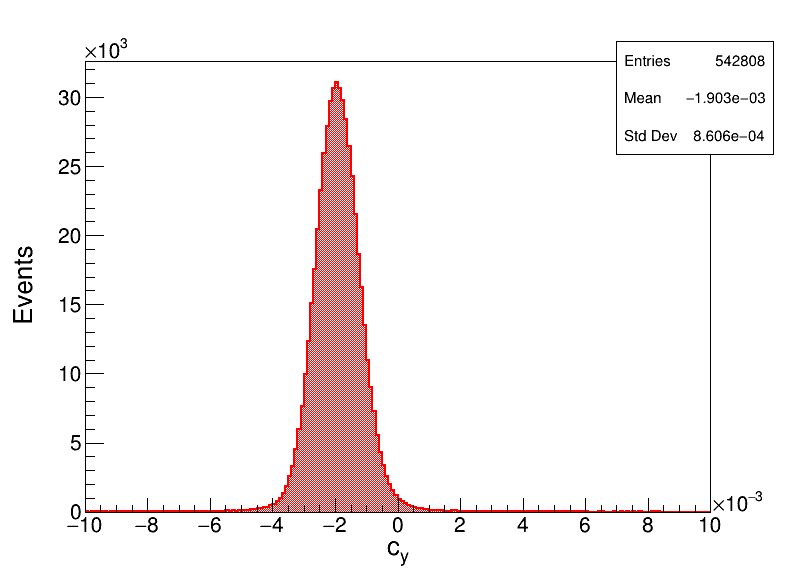}
\includegraphics[width=0.48\textwidth]{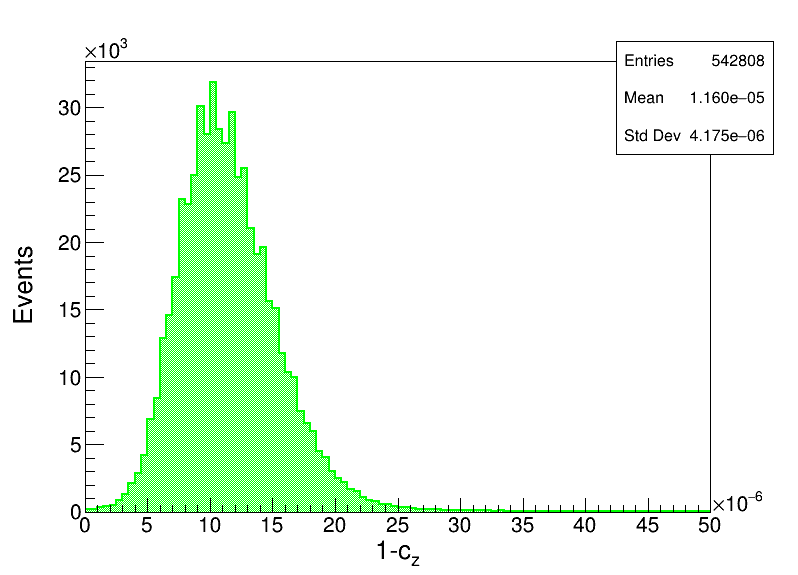}
\end{center}
\caption{Distributions of the direction cosines of the electron tracks in the silicon detector and in the beam telescope in the alignment run.}
\label{fig:dircosines} 
\end{figure}

Fig.~\ref{fig:dircosines} shows the distributions of the direction cosines of the electron tracks in the alignment run. We see that the average values of the direction cosines $c_x$ and $c_y$ are slightly different from zero. This result implies that the $z$-axis of our reference frame is not perfectly aligned with the direction of the beam. The tilt angle can be estimated from the average value of $c_z$, and is of about $5\unit{mrad}$. Finally, from the values of the RMS of the distributions of $c_x$ and $c_y$ we can deduce that the beam divergence is of about $1\unit{mrad}$ in both the $x$ and $y$ directions.

\subsection{Data selection and analysis}

As discussed in Sec.~\ref{sec:setup}, several runs in different configurations have been taken, by changing the beam composition and momentum, the radiator and its distance from the silicon pixel detector.

In each of these runs we have selected events with at least one cluster in the silicon pixel sensor and at least 3 clusters in different detectors of the beam telescope. This choice is motivated by the need of identifying, among the clusters in the silicon sensor, the one originated by the ionization energy deposit of the beam particle and those eventually originated by the absorption of TR X-rays produced in the upstream radiator.

\begin{figure}[!t]
\begin{center}
\includegraphics[width=0.8\textwidth]{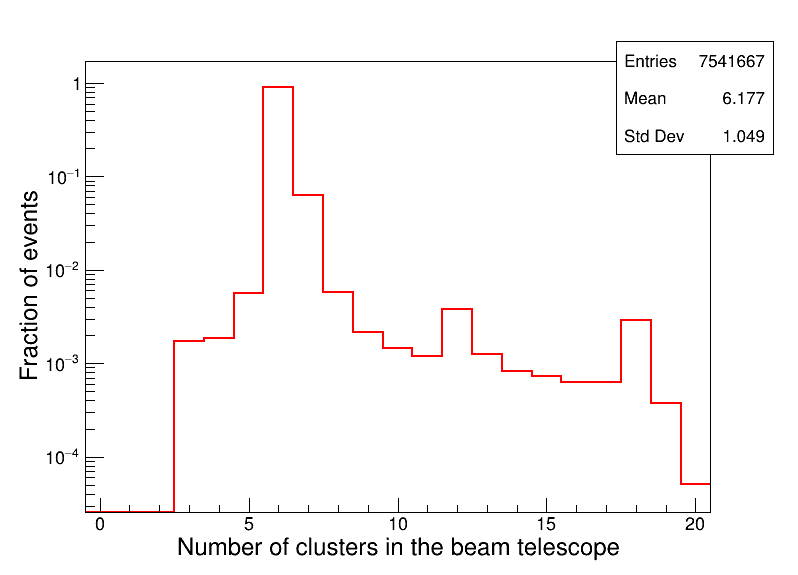}
\end{center}
\caption{Distribution of the total number of clusters in the beam telescope for all the runs performed with electrons crossing the EXTRA radiator, placed at a distance of $88.9\unit{cm}$ from the silicon pixel sensor.}
\label{fig:clusters_telescope} 
\end{figure}

Fig.~\ref{fig:clusters_telescope} shows the distribution of the total number of clusters in the detectors of the beam telescope for all the runs performed with electrons crossing the EXTRA radiator, which was placed at a distance of $88.9\unit{cm}$ from the silicon pixel sensor. As expected, the distribution is peaked at 6 clusters, corresponding to clean electron tracks, yielding one cluster in each detector. Events with less than 6 clusters can be originated from inefficiencies of some detectors in the beam telescope or from beam particles which do not cross all the telescope planes. Events with more than 6 clusters can be originated from delta rays accompanying the primary electron track or from TR X-rays passing through the upstream silicon sensor and being absorbed in any detector of the silicon telescope. We also see two peaks, at 12 and 18 clusters respectively, which include less than $1\%$ of the total number of events, and which likely correspond to double and triple electron tracks.

The clusters in the detectors of the beam telescope are used to reconstruct the tracks of the beam particles in the telescope. To select events with single electron (positron) tracks, we require less than $10$ clusters in the beam telescope. Candidate tracks are built by selecting all the possible cluster combinations with only one cluster per plane of the telescope. The clusters of each candidate track are then fitted with a straight line and the $\chi^{2}$ of the fit is evaluated. The track with the best $\chi^{2}$ is then selected. 

\begin{figure}[!t]
\begin{center}
\includegraphics[width=0.8\textwidth]{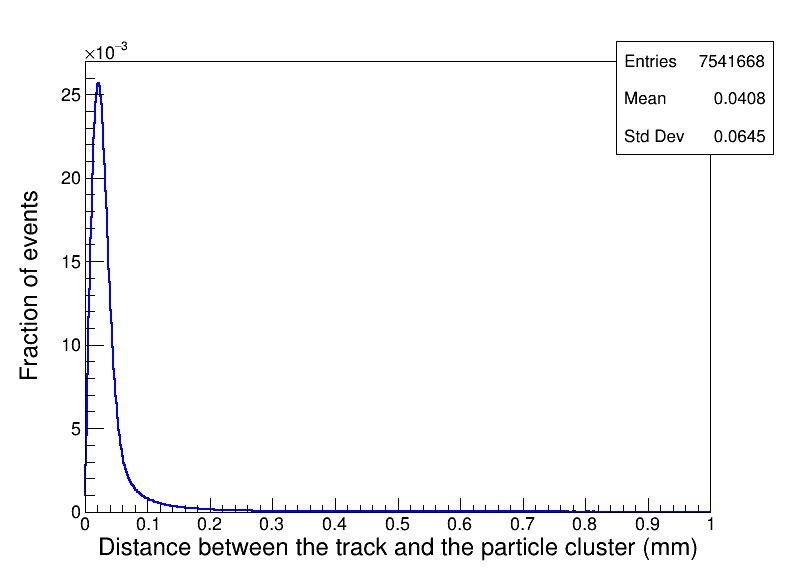}
\end{center}
\caption{Distribution of the distances of the "particle clusters" from the track in the silicon sensor for the runs performed with electrons crossing the EXTRA radiator, placed at a distance of $88.9\unit{cm}$ from the silicon pixel sensor.}
\label{fig:distances} 
\end{figure}

Once the track of the radiating particle in the beam telescope is reconstructed, we evaluate the coordinates $(x_{track},y_{track})$ of its intersection with the upstream silicon pixel sensor. Then, if more than one cluster is found in the sensor, the cluster nearest to the track is associated to the particle ("particle cluster"), while other clusters are associated to possible TR X-rays ("X-ray clusters"). Clearly, if only one cluster is found in the silicon pixel sensor, it is associated to the particle and no X-rays are detected. Fig.~\ref{fig:distances} shows the distribution of the distances of the particle clusters in the silicon pixel sensor from the track reconstructed in the beam telescope obtained in the runs performed with electrons crossing the EXTRA radiator. We see that $94\%$ of the particle clusters are within $100\unit{\mu m}$ from the track and only $0.1\%$ of them are at distances above $500 \unit{\mu m}$.

\begin{figure}[!h]
    \centering
    \includegraphics[width=0.85\textwidth,height=0.28\textheight]{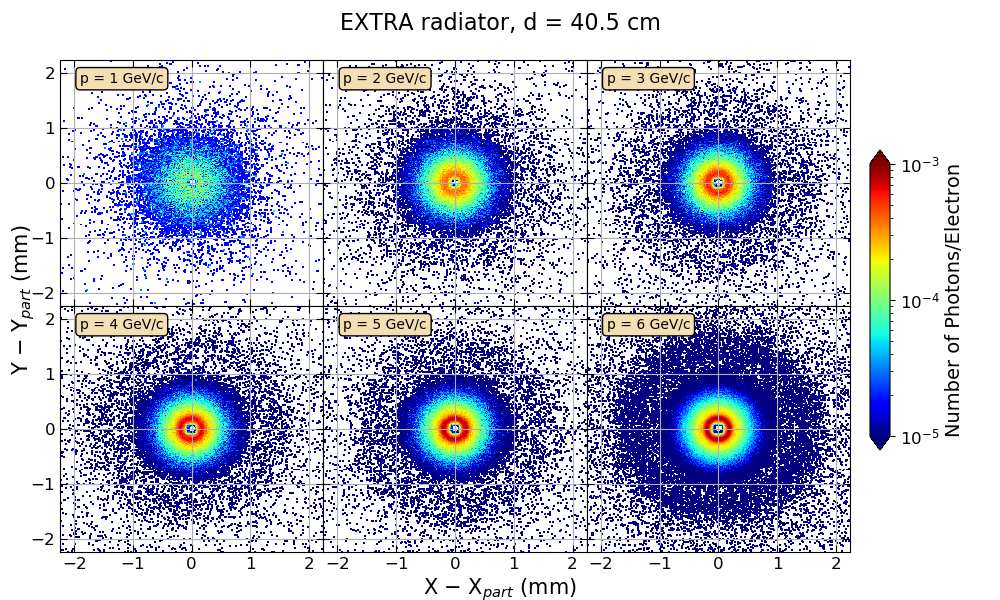}
    \includegraphics[width=0.85\textwidth,height=0.28\textheight]{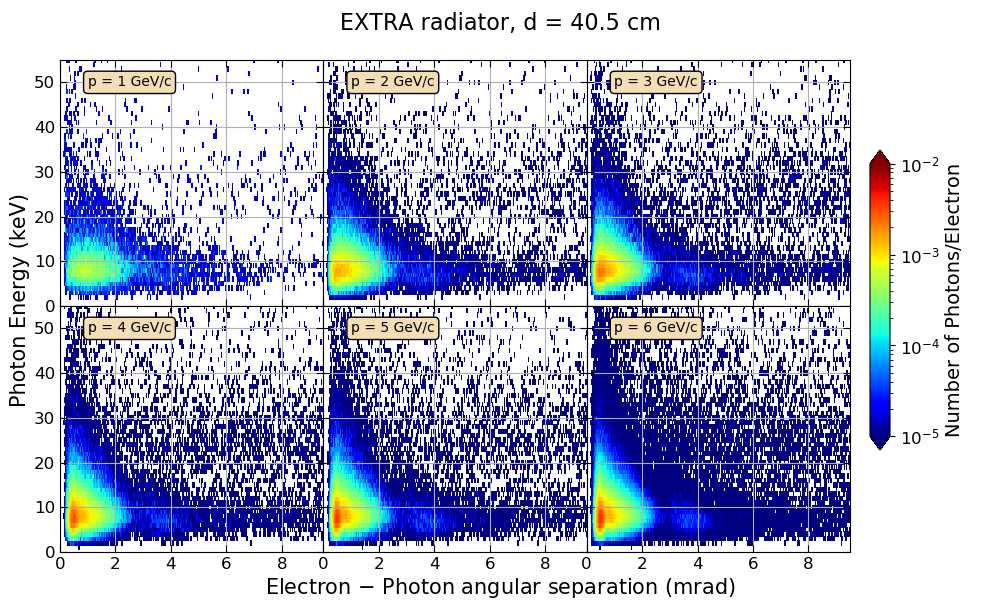}
    \includegraphics[width=0.85\textwidth,height=0.28\textheight]{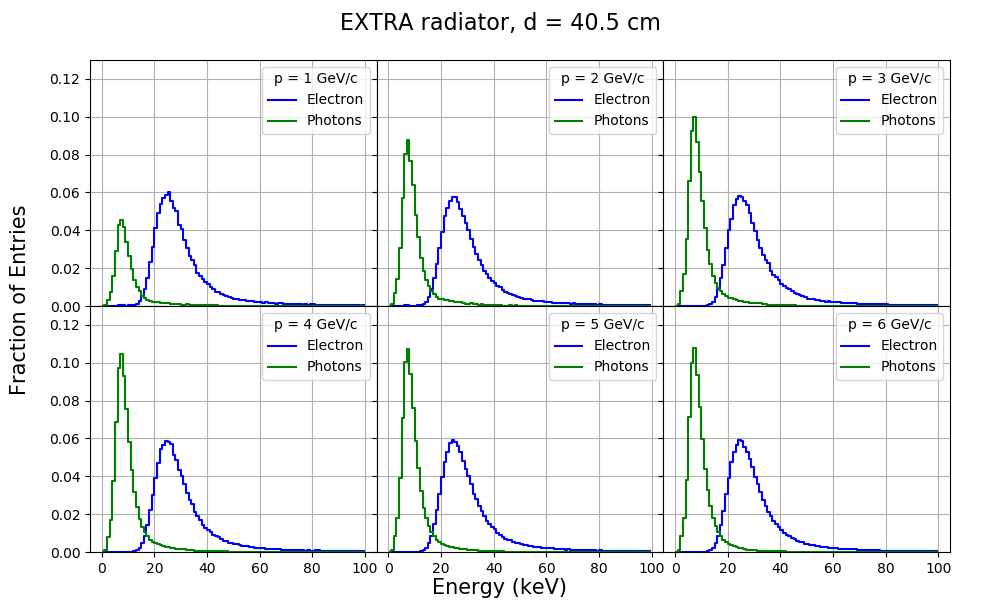}
    \caption{Summary of the results obtained in the runs with the EXTRA radiator at $40.5\unit{cm}$ from the Timepix3 detector. Top panel: distribution of the relative positions of the TR photons (X-ray clusters) with respect to the electrons (particle clusters); middle panel: distribution of X-ray energies as a function of their angular separation from the electrons; bottom panel: electron and X-ray energy distributions.}
    \label{fig:Extra_40}
\end{figure}

\begin{figure}[!ht]
    \centering
    \includegraphics[width=0.85\textwidth,height=0.28\textheight]{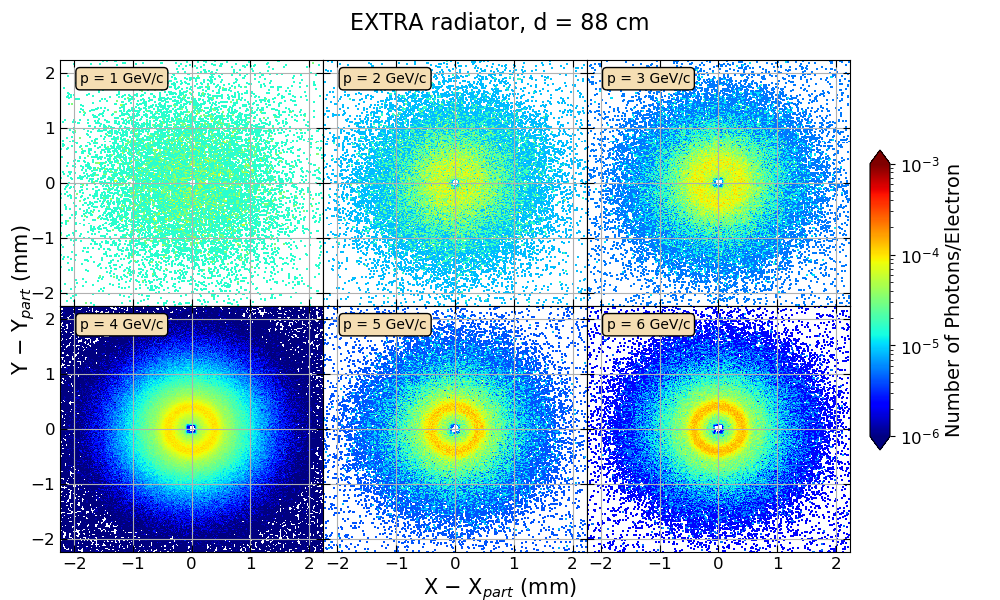}
    \includegraphics[width=0.85\textwidth,height=0.28\textheight]{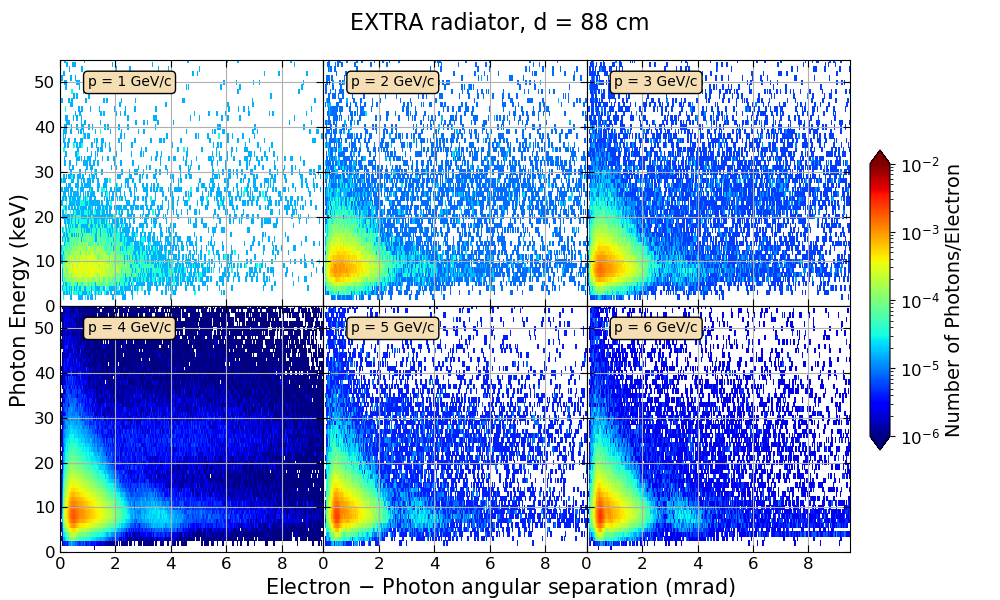}
    \includegraphics[width=0.85\textwidth,height=0.28\textheight]{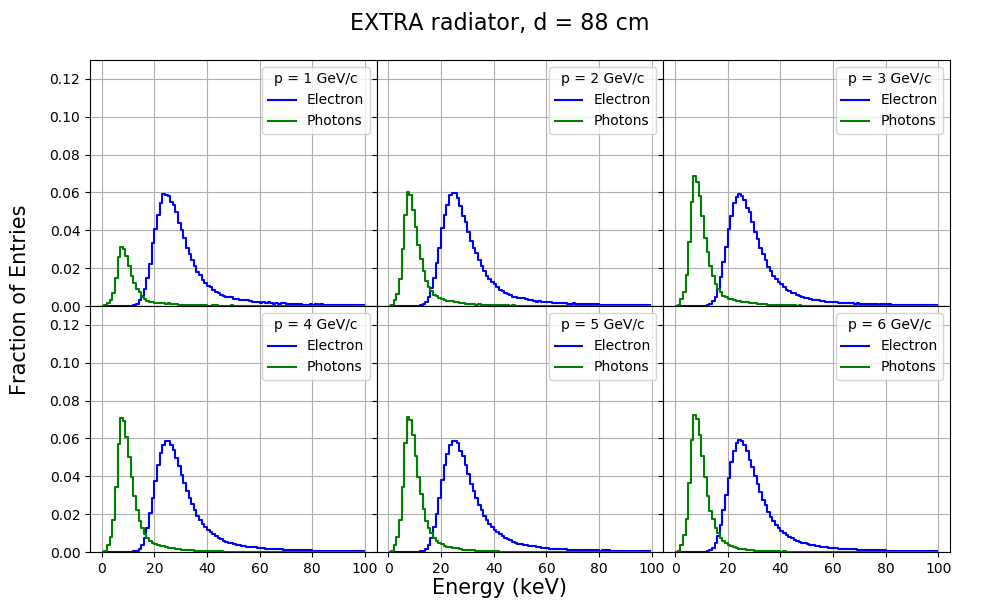}
    \caption{Summary of the results obtained in the runs with the EXTRA radiator at $88\unit{cm}$ from the Timepix3 detector. Top panel: distribution of the relative positions of the TR photons (X-ray clusters) with respect to the electrons (particle clusters); middle panel: distribution of X-ray energies as a function of their angular separation from the electrons; bottom panel: electron and X-ray energy distributions.}
    \label{fig:Extra_88}
\end{figure}

\begin{figure}[!ht]
    \centering
    \includegraphics[width=0.85\textwidth,height=0.28\textheight]{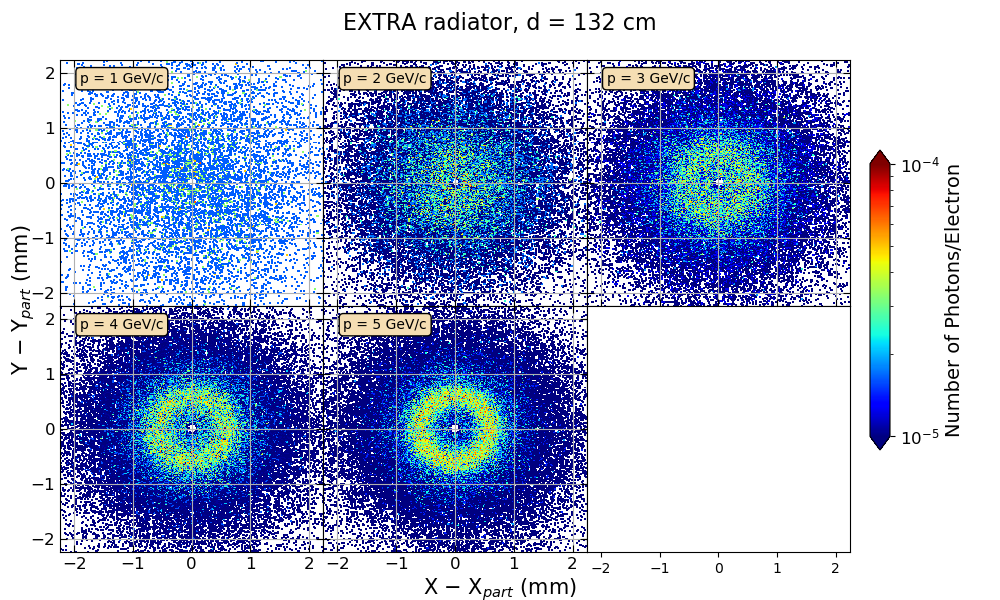}
    \includegraphics[width=0.85\textwidth,height=0.28\textheight]{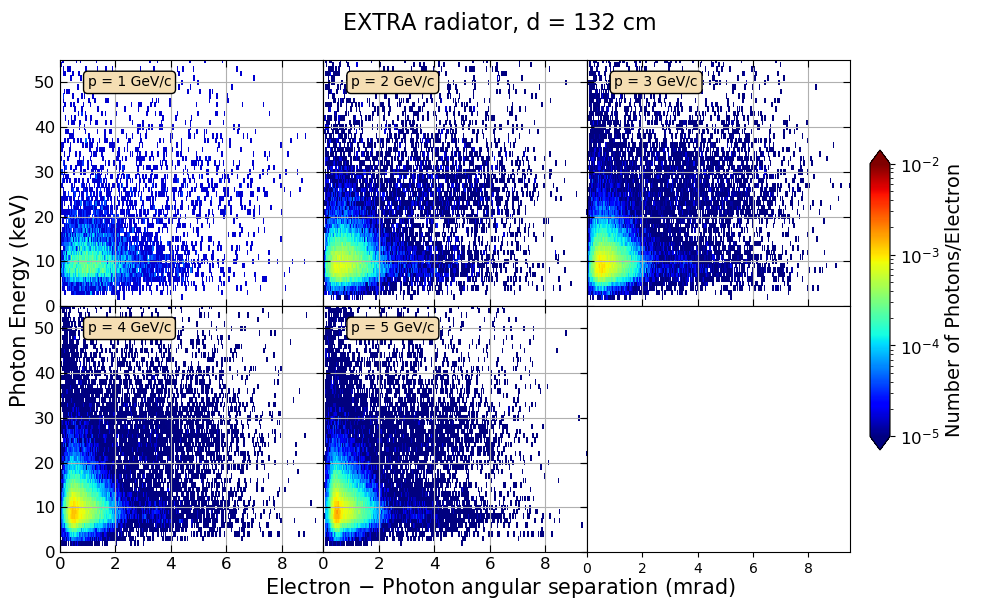}
    \includegraphics[width=0.85\textwidth,height=0.28\textheight]{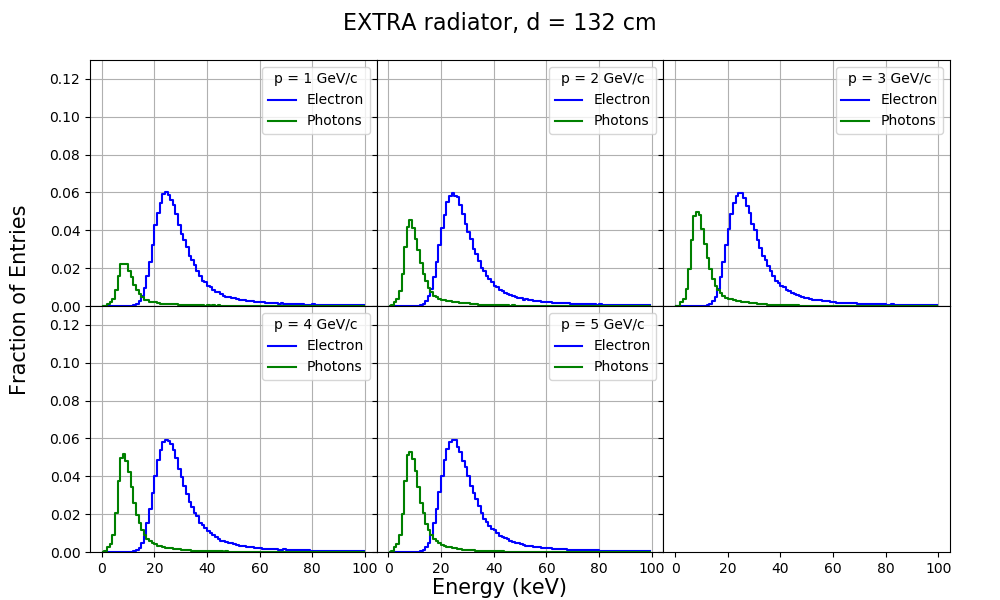}
    \caption{Summary of the results obtained in the runs with the EXTRA radiator at $132\unit{cm}$ from the Timepix3 detector. Top panel: distribution of the relative positions of the TR photons (X-ray clusters) with respect to the electrons (particle clusters); middle panel: distribution of X-ray energies as a function of their angular separation from the electrons; bottom panel: electron and X-ray energy distributions.}
    \label{fig:Extra_132}
\end{figure}

\section{Results}
\label{sec:result}

In Figs.~\ref{fig:Extra_40},~\ref{fig:Extra_88} and~\ref{fig:Extra_132} the results obtained in the runs with the EXTRA radiator are summarized. The plots in each figure correspond to the configurations with the silicon detector placed at the distances of $40.5 \unit{cm}$, $88 \unit{cm}$ and $132 \unit{cm}$ from the radiator respectively. The plots are built selecting events with the particle cluster inside a square of a $3\times 3 \unit{mm^2}$ area, in the centre of the Timepix3 detector. All the distributions shown in the above plots are normalized to the total number of selected events. 

The top panels of each figure show the distributions of the relative positions of the TR X-rays (evaluated from the X-ray clusters) with respect to the radiating electron (evaluated from the particle cluster). As expected, TR photons tend to accumulate in rings centered on the position of the radiating particle and the number of photons per electron increases with the beam momentum (and consequently with the Lorentz factor of the radiating particles). 

The central panels show the distributions of the TR X-ray energies as a function of their angular separation from the radiating particle. Most X-rays are emitted at angles $\theta \lesssim 2\unit{mrad}$ from the radiating particle, with energies peaked at energies $<10 \unit{keV}$. A second peak of X-rays emitted at angles $\sim 3.5\unit{mrad}$ and with the same energies as the first peak can also be seen, and it becomes more evident as the beam momentum increases. 

Finally, the bottom panels show the energy distributions of the absorbed TR X-rays compared with the distributions of the energies deposited by the parent electrons in the Timepix3 detector. As discussed in Sec.~\ref{sec:clustering}, the energy losses of the electrons follow Landau distributions with a most probable value of $25.4\unit{keV}$, while X-ray energies are peaked at less than $10\unit{keV}$. We see that the area of the X-ray energy spectra increases with increasing electron momentum. This behaviour is expected since the spectra are normalized to the total number of electrons and the TR yield increases with the Lorentz factor of the radiating particle.

\begin{figure}[!t]
    \centering
    \includegraphics[width=1\textwidth,height=0.45\textheight]{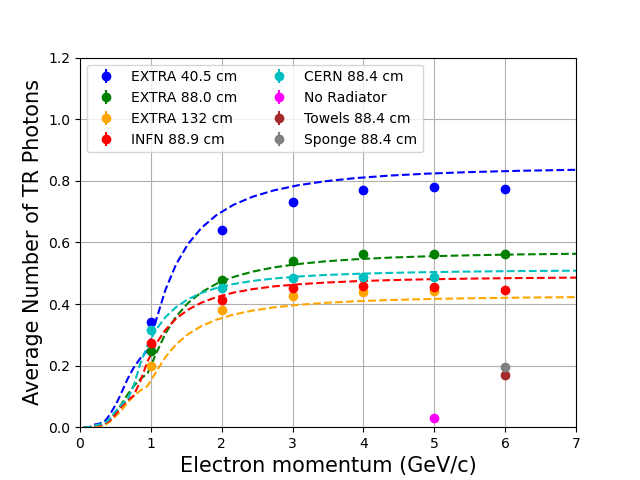}
    \caption{Average number of TR photons as a function of electron beam momentum for the three radiator types and for the different distances from the Timepix3 detector. The dashed lines show the predictions for the different configurations. The results obtained in a run without radiator and in two runs with  "dummy" radiators are also shown.}
    \label{fig:TRall}
\end{figure}

A summary of the results obtained in all the configurations explored is shown in Fig.~\ref{fig:TRall}. The average number of detected TR X-rays per electron is shown as a function of the beam momentum. We see that for all configurations the number of detected photons increases with the beam momentum and saturates above $4\unit{GeV/c}$. This behavior is expected, since the threshold Lorentz factor for all radiators is $\gamma_{thr} \simeq 10^{3}$ and the saturation Lorentz factors are in the range $4 \div 5 \times 10^{3}$. Comparing the results obtained with the EXTRA radiator in the different configurations we see that the average number of detected TR X-rays decreases when the radiator-detector distance is increased. The increase of the distance causes an increase of the X-ray absorption in the air gap between the radiator and the detector, which is not compensated by the lower minimum detectable angle between the photons and the radiating particles. We remark here that the results shown in Fig.~\ref{fig:TRall} referred to the CERN radiator have been obtained from a joint analysis of the data samples collected with both the electron and positron beams (see Tab.~\ref{tab:beam}). This choice is motivated by the fact that the separate analyses of the electron and positron data samples yield the same results. This feature was expected, since the properties of TR are independent of the sign of the charge of the radiating particle.

The experimental results shown in Fig.~\ref{fig:TRall} are compared with the predictions obtained by folding the TR yield, evaluated with the theoretical formulae for regular radiators~\cite{Artru:1975xp,Fabjan:1975rh} with the X-ray absorption probabilities in the air gap between the radiator and the Timepix3 detector and in the silicon layer of the detector. The theoretical curves seem to be in a reasonable agreement with the experimental results.

Finally, we have performed some control runs to check our results. A run with $5 \unit{GeV/c}$ electrons without any radiator was performed to evaluate the possible contamination to the detected TR signal from bremsstrahlung photons produced in the upstream materials and accompanying the beam particles and the possible contamination from noisy pixels. In this run we found about $0.03$ X-rays per electron; in addition, since all X-ray clusters are found very close to the particle cluster, the contamination from noisy pixels can be considered negligible. We also performed two additional runs with $6 \unit{GeV/c}$ electrons, in which we replaced the radiator with some "dummy" radiators: in particular, we used a set of paper towels, which were arranged in a stack simulating a regular radiator, and a piece of sponge, which simulates an irregular radiator\footnote{Irregular radiators made of foams or fiber mats are sometimes used in TRDs.}. In both cases we observed a TR signal of about $0.17$ X-rays per electron, indicating that TR can be generated also at everyday objects. At the same time it should be noted that the design and quality of a radiator is crucial to optimise the photon yield and thus the signal in corresponding detector systems.

\section{Conclusions}
\label{sec:conclu}

In the framework of the BL4S competition we have designed and implemented an experiment to measure the TR emitted by fast electrons and positrons crossing different kind of radiators. The measurement has been performed at the DESY II Test Beam Facility area TB21, using electrons and positron beams with momenta up to $6\unit{GeV/c}$. We have measured the energy spectra and the angular distribution of the TR X-rays using a $100 \unit{\mu m}$ thick pixel silicon detector, with a pitch of $55\unit{\mu m}$. The experimental results are well reproduced by the theoretical curves obtained from standard TR models.

BL4S has offered the students the chance to be actively involved in all the aspects of an experimental research: during the preparation of the proposal, they have learned how to design an experimental setup, optimizing the detectors available for the measurement; after their proposal was selected, they have been involved in the design and in the assembly of their own radiator; then, at DESY, they had the chance to run a real beam test; finally, they have taken part to the analysis of the data collected in the test. However, the most important educational result of this experience is that the students learned how to apply the scientific approach not only in the field of research, but also to the solution of everyday life challenges.

\section*{Acknowledgments}

The measurements leading to these results have been performed at the Test Beam Facility at DESY Hamburg (Germany), a member of the Helmholtz Association (HGF). 
The Timepix3 assembly used for the measurements within this publication is provided by the CLICdp group at CERN.

The members of the EXTRA team thank the CERN and DESY support scientists, the beamline scientists, the volunteers and the BL4S organisers who helped them during the preparation and the implementation of their experiments. All the scientists involved in the competition dedicated a lot of their time to answer all the questions the students had, giving them precious advises for their future career. The team was really pleased to find such wonderful people, who showed them what unconditional love for science really means. 

A big thank to the Teomizli team from Mexico, the other winning team of the BL4S 2021. Meeting peers from the other side of the world and work with them as a unique team of scientists has been an enriching opportunity.

Beamline for Schools is an education and outreach project funded by the CERN \& Society Foundation and supported by individual donors, foundations and companies. In 2021, the project was funded by the Wilhelm and Else Heraeus Foundation. Additional contributions have been received from the Arconic Foundation, Amgen Switzerland AG, and the Ernest Solvay Fund managed by the King Baudouin Foundation. 

The EXTRA team also acknowledges financial support from CERN and DESY for their participation to the beam test campaign.

The EXTRA team thanks B.~Fanti, I.~Iusco, D.~Ricchiuti and all the personnel of the Liceo Scientifico ``A.~Scacchi'' for their support to the project activities.

\bibliographystyle{unsrt}
\bibliography{biblio.bib}{}

\end{document}